# Effective dislocation lines in continuously dislocated crystals
## I. Material anholonomity

ANDRZEJ TRZĘSOWSKI

**Institute of Fundamental Technological Research**
**Polish Academy of Sciences**
Świętokrzyska 21, 00-049 Warsaw, Poland
e-mail: atrzes@ippt.gov.pl or atrzes@wp.pl

A continuous geometric description of Bravais monocrystals with many dislocations and secondary point defects created by the distribution of these dislocations is proposed. Namely, it is distinguished, basing oneself on Kondo and Kröner's Gedanken Experiments for dislocated bodies, an anholonomic triad of linearly independent vector fields. The triad defines local crystallographic directions of the defective crystal as well as a continuous counterpart of the Burgers vector for single dislocations. Next, the influence of secondary point defects on the distribution of many dislocations is modeled by treating these local crystallographic directions as well as Burgers circuits as those located in such a Riemannian material space that becomes an Euclidean 3-manifold when dislocations are absent. Some consequences of this approach are discussed.

### 1. Introduction

Let us take 1 mm as a macroscopic observation level scale and let 1 Å (the diameter of the hydrogen atom in the ground state) define the atomic-size observation level scale. It is known that for usual well-annealed pure metals, the mean distance between dislocations is of the order of $1\,\mu m$ $(1\,\mu m = 10^{-3}\,mm)$, and that the crystal with many dislocations can be treated, on a mesoscopic observation level scale that lies e.g. in the range of 10-100 nm ($1 \text{ nm} = 10^{-3} \mu\text{m} = 10 \text{ Å}$), as a part of an ideal crystal [1]. On the other hand, if the macroscopic properties of a crystalline solid with many dislocations are considered, a *continuous limit approximation* can be defined by means of the condition that, at each point of the body, a *characteristic mesoscopic length*, say of the order of 10-100 nm, can be approximately replaced with the infinitesimal length. Consequently, a monocrystal with many dislocations can be considered as a such *locally homogeneous* continuous body that retains locally the most characteristic properties of the original crystal, namely the existence of crystallographic directions at each point, the lattice rotational symmetries (the lattice translational symmetries are lost in a continuous limit) and the mass per unit volume. Note that in this continuous limit, the content of defects (e.g. the so-called *scalar density of dislocations* defined as the length of all dislocation lines included in the volume unit) remains unchanged. Therefore, although the global long range-order of crystals is lost in the presence of dislocations, nevertheless their local long-range order still exists [2]. It is represented by the *object of material anholonomity* of the continuized dislocated crystal (Sections 2 and 4). We restrict our investigations to the *Bravais crystals*, because these crystals have the smallest amount of different defect types, but enough to study the general principles [3].

The appearance of dislocations generates a bend of originally straight lattice lines. For example, the *lattice lines* in a continuized dislocated Bravais crystal form a system of three independent congruences of curves and tangents to these curves define *local crystallographic directions* of this crystal (Section 2). Planes spanned by two local crystallographic directions are *local crystal planes*. In general, none of these congruenences is normal (that is the curves of the congruence are not orthogo-nal trajectories of a family of surfaces). If a *crystallographic congruence* is normal and its curves are orthogonal to local crystal planes everywhere, then the curves are orthogonal trajectories of *crystal surfaces* of the continuously dislocated Bravais crystal (Section 6). If, additionally, local crystal planes of a crystal surface are the virtually *local slip planes*, the crystal surface is virtually a slip surface along which a curved dislocation line can move. The crystal surface is called then a *glide surface*. Particularly, in the case of *single glide* crystal planes originally parallel and normal to a lattice direction pass into slip surfaces without local stretchings. In this case the glide surface is flat [4].

It is known that the occurrence of many dislocations in a crystalline solid is accompanied with the appearance of *secondary point defects* (vacancies and self-interstitials) created by the distribution of dislocations. It is, for example, due to intersections of dislocation lines: point defects can appear in crossover points of edge dislocation lines or when two parallel dislocation lines are joinined together [5]. In a real crystal the existence of point defects influences the distance between lattice points (e.g. [2], [3] and [6]). On the other hand, as a crystal with many dislocations reveals the *short-range order* (Section 2), dislocations have no influence on local metric properties of a crystal structure. It means that we are dealing with the locally Euclidean geometry of defective crystals. Therefore, the influence of secondary point defects on the metric properties of a continuized dislocated Bravais crystal can be represented by the such Riemannian *internal length measurement* that reduces to the Euclidean one (induced from the environment of the crystal) when dislocations are absent (Section 3). The influence of secondary point defects on lattice lines, crystal surfaces and dislocation densities (tensorial as well as scalar) can be described, in a continuous limit, by means of treatment of these lines, surfaces and Burgers circuits as those located in the *Riemannian material space* (Sections 4-6). Since the material anholonomy can be represented by a plastic deformation of the defective Bravais crystal (Sections 2 and 3) and the work done in plastic deformations more than 90% goes into heat [7], it seems reasonable to discuss the problem of thermal distortions of the above-mentioned internal length measurement (Section 7).

## 2. Bravais moving frame

Assume that a stress-free crystalline solid is loaded by boundary tractions in the elastic regime. The occurrence of crystalline structure defects can be recognized by the fact that unloading does not take the body back to its original configuration. The unloaded state will thus contain a residual stress field. On the other hand, we assume that the stored energy is only due to elastic deformations and clearly residual stresses cannot be captured by a deformation gradient because these would model a body that unloads completely. In the case of dislocated bodies we can characterize deformations of that unloaded state based on an assumption that the distorted lattice is uniquely defined everywhere. Namely, following the so-called *Kondo cutting-relaxation procedure*, one images removing of an infinitely small part of a dislocated crystalline body and allowing it to relax (by removing all boundary tractions) up to an unstrained state called a *natural state*. The discrete material structure of the natural state coincides with a perfect lattice and we can use the difference between these states and the deformed state as a measure of the stored elastic energy. Let us consider, in order to describe this measure explicitly, a reference configuration $B \subset E^3$ of the body (being an open and connected subset of the Euclidean point space $E^3$ and frequently identified with the body itself). Let $(U, X)$, $X = (X^A)$ be a curvilinear coordinate system on $B$ (called Lagrange coordinates) and let $x = (x^i)$ be a curvilinear coordinate system on $E^3$ (called Eulerian coordinates) such that $[x^i] = [X^A] = \text{cm}$. Let us denote $p = p(X) \in B$ iff $X = (X^A) = X(p) \in \mathbb{R}^3$ and let $dX^A$ denote an infinitesimal distance localized at the point $p$ in the $\mathbf{c}_A(p)$-direction tangent to the coordinate curve $\sigma_A$ (see Appendix). If $\delta x^i$ denotes the corresponding infinitesimal distance localized in a deformed state $B_\lambda = \lambda(B)$, where $\lambda : B \to E^3$ is a diffeomorphism (a deformation), and $\delta \xi^a$ is the same relaxed material line element, then the relations [2]

$$\delta x^i = F^i{}_A(X) dX^A = B^i{}_a(X) \delta \xi^a,$$
(2.1)
$$\delta \xi^a = P^a{}_A(X) dX^A,$$

where

$$F^i{}_A(X) = \lambda^i{}_{,A}(X),$$
(2.2)
$$\lambda^i = x^i \circ \lambda \circ X^{-1} : X(U) \to \mathbb{R}^3,$$

define the *distortions* of the body: total $(F^i{}_A)$, plastic $(P^a{}_A)$ and elastic $(B^i{}_a)$.

The *Kondo Gedanken Experiment* means the repetition of Kondo cutting-relaxation procedure for many small elements of the body. We obtain then an amorphous collection of small line elements of a crystalline solid with a perfect lattice, which are translated and rotated with respect to one another and therefore, to mesh to form a homogeneous Euclidean material continuum. It means, among other things, the discrepancy of relaxed material line elements $\delta \xi^a = \delta \xi^a (X(p))$, $p \in B$, located at the body points. If we will define the moving coframe $\Phi^* = (E^a; a = 1, 2, 3)$, by

$$(2.3) \quad \mathrm{E}^a(p(X)) = \overset{a}{e}_A(X) d\mathrm{X}^A_{p(X)} \quad \text{for} \quad X = \mathrm{X}(p) \in \mathbb{R}^3,$$
$$\overset{a}{e}_A(X) = P^a{}_A(X), \quad p = p(X) \in B, \quad [\mathrm{E}^a] = \text{cm},$$

and identify the infinitesimal material line element $dX^A$ with the covector $dX^A_{p(X)}$ at a point $p(X) \in B$, then the *translational discrepancy* of those material line elements can be represented by the condition that $\Phi^*$ is nonintegrable, i.e. by the condition that, for at least one 1-form $\mathrm{E}^a$, should be

$$(2.4) \quad \tau^a = d\mathrm{E}^a \neq 0.$$

The ordered set $\Phi^*$ of 1-forms $\mathrm{E}^a$ represents then a translational discrepancy of the family $\{\delta\xi^a(\mathrm{X}(p)), \ p \in B\}$ of relaxed material line elements. This representation of the translational discrepancy can be treated as a continuous limit neglecting the finiteness of the lattice spacing. We can think, for example, of some limiting process in which lattice constants of a Bravais lattice decrease more and more, but the lattice rotational symmetries as well as the mass per unit volume and the content of defects remain unchanged. The resulting body, called by Kröner a *continuized crystal* [3], retains locally the most characteristic properties of the original crystal (Section 1).

Let us consider a moving frame $\Phi = (\mathbf{E}_a; \ a = 1, 2, 3)$ of base vector fields, tangent to the continuized Bravais crystal $B$ and parallel to its local crystallographic directions, as the one defining relaxed material line elements of the Kondo Gedanken Experiment according to Eq. (2.3) and the following duality condition:

$$(2.5) \quad \mathbf{E}_a(p) = \overset{A}{e}{}_a(\mathrm{X}(p)) \partial_{A|p}, \quad p \in B,$$
$$\mathrm{E}^a(\mathbf{E}_b) \equiv \langle \mathrm{E}^a, \ \mathbf{E}_b \rangle = \overset{a}{e}_A \overset{A}{e}{}_b = \delta^a_b, \quad [\mathbf{E}_a] = \text{cm}^{-1}.$$

We will call $\Phi$ a *Bravais moving frame* [8]. Note that from the *Kröner's Gedanken Experiment* (i.e. the continuization procedure) follows the existence of the following *local rotational uncertainty* to select the Bravais moving frame [2]:

(2.6)
$$\mathbf{E}_a(p) \to \mathbf{E}'_a(p) = Q^b{}_a(p)\mathbf{E}_b(p),$$
$$Q = (Q^a{}_b): \quad B \to G \subset SO(3),$$

where SO(3) denotes the proper orthogonal group in $\mathbb{R}^3$ and G is the group of point symmetries of an ideal Bravais lattice defining a discrete crystalline structure of the body under consideration. A pair ($\Phi$, G) represents the *short-range order* of the dislocated crystalline solid treated as a locally homogeneous body [2].

The elastic distortion $B^k{}_a$ of Eq. (2.1) is a measure of the stored elastic energy and defines the following *elastic transformation* of the Bravais moving coframe $\Phi^*$:

(2.7)
$$F^k(p) = B^k{}_a(X(p))E^a(p) = F^k{}_A(X(p))dX^A_p,$$

where Eqs. (2.2) and (2.3) were taken into account. If $\lambda: B \to E^3$ is a deformation, then the *total deformation tensor* $\mathbf{F}(p)$ fulfils for any $\mathbf{u} \in E^3$ the condition:

(2.8)
$$d\lambda_p(\mathbf{u}) = \mathbf{F}(p)\mathbf{u}, \qquad \mathbf{F}(p) = \partial_{k|\lambda(p)} \otimes F^k(p).$$

Let us consider a continuous solid body $B \subset E^3$ with its material structure defined by the continuous limit approximation of a Bravais crystal with many dislocations. A distinguished basis $\Phi = (\mathbf{E}_a; \ a = 1,2,3)$ of the linear module W(*B*) (see Appendix), can be identified with a *Bravais moving* frame that defines a system of three independent congruences of *lattice* lines of the continuized crystal (Section 1). The condition that the bend of lattice lines due to dislocations (Section 1) is not generated by a global deformation of the body means that the *object of material anholonomity* $C_{ab}^c \in C^\infty(B)$ (called in differential geometry an object of anholonomity of the manifold $B$ [9]) defined by Eq. (A.7) and by

$$[\mathbf{E}_a, \mathbf{E}_b] = C_{ab}^c \mathbf{E}_c, \tag{2.9}$$

does not vanish. Next, let us define a tensorial representation $\mathbf{S}[\Phi]$ of anholonomity of the base field $\Phi = (\mathbf{E}_a)$ on $B$. Namely, if $\Phi^* = (E^a)$ is the Bravais moving coframe dual to $\Phi$, then we define:

$$\mathbf{S}[\Phi] = dE^a \otimes \mathbf{E}_a = S_{ab}{}^c E^a \otimes E^b \otimes \mathbf{E}_c,$$
$$S_{ab}^c \in C^\infty(B), \qquad [S_{ab}{}^c] = cm^{-1}. \tag{2.10}$$

It can be shown that [10]

$$S_{ab}{}^c = -\frac{1}{2} C_{ab}^c, \tag{2.11}$$

and $\mathbf{S}[\Phi]$ characterizes the existence of many dislocations in this sense that

$$\mathbf{S}[\Phi] = \mathbf{0} \quad \text{iff} \quad \mathbf{E}_a = \partial/\partial \xi^a, \quad a = 1, 2, 3, \tag{2.12}$$

where $\xi = (\xi^a)$ is a coordinate system on $B$. Thus, the field $\mathbf{S}[\Phi]$ can be interpreted as a nondimensional measure of the *long-range distortion* of the continuized dislocated Bravais crystal due to a bend of originally straight lattice lines (Section 1) [2].

The long-range distortion of a dislocated Bravais crystal can be quantitatively measured by the so-called *Burgers vector* corresponding to a closed contour $\gamma$ in this crystal (e.g. [11] and [12]). Let us consider, in order to formulate the definition of a continuous counterpart of this vector, a family $\lambda = \{\lambda_p : U_p \to E^3, \ p \in U_p \subset B\}$ of local diffeomorphisms of the body $B \subset E^3$ defined on an open covering $\{U_p, \ p \in B\}$ of $B$ [13]. Let us assume that for arbitrary $p \in B$ and for any Lagrange coordinate system coordinate system $X: U_p \to \mathbb{R}^3$, the following relations hold:

$$\forall q \in U_p, \quad d\lambda_{p,q}: \mathrm{T}_q(B) \to \boldsymbol{E}^3,$$
(2.13)
$$\lambda_p(p) = O \in \boldsymbol{E}^3, \quad d\lambda_p(\mathbf{E}_a)(p) = d\lambda_{p,p}(\mathbf{E}_a(p)) = \mathbf{C}_a \in \boldsymbol{E}^3,$$

where $O \in \boldsymbol{E}^3$ is a distinguished point, $\Phi = (\mathbf{E}_a; a = 1, 2, 3)$ is a Bravais moving frame, $C = (\mathbf{C}_a; a = 1, 2, 3)$ is an orthonormal Cartesian basis of the Euclidean vector space $\boldsymbol{E}^3$ of translations in $E^3$, $\mathrm{T}_q(B)$ denotes the vector space of all vectors tangent to $B$ at a point $q \in U_p$ and spanned by the vectorial basis $\Phi_q = (\mathbf{E}_a(q))$, and Eqs. (2.7) and (2.8) were taken into account. Let $\gamma$ be a closed contour in $B$ passing through the point $p \in B$ and with its tangent vector field $\dot{\boldsymbol{\gamma}}$:

$$\gamma: \langle \alpha, \beta \rangle \to B, \quad \gamma(\alpha) = \gamma(\beta) = p,$$
(2.14)
$$\dot{\boldsymbol{\gamma}}(t) = \dot{\gamma}^a(t)\mathbf{E}_a(\gamma(t)) \in \mathrm{T}_{\gamma(t)}(B), \quad t \in (\alpha, \beta).$$

If $\gamma(\langle \alpha, \beta \rangle) \subset U_p$, then the vector field

(2.15)
$$\forall t \in (\alpha, \beta), \quad \dot{\boldsymbol{\gamma}}_\lambda(t) = d\lambda_{p,\gamma(t)}(\dot{\boldsymbol{\gamma}}(t)) = \dot{\gamma}^a(t)\mathbf{C}_a \in \boldsymbol{E}^3,$$

afford possibilities for the definition of a curve $\gamma_\lambda$ in $E^3$:

(2.16)
$$\gamma_\lambda: \langle \alpha, \beta \rangle \to E^3, \quad \overrightarrow{O\gamma_\lambda(t)} = \int_\alpha^t \dot{\boldsymbol{\gamma}}_\lambda(s)ds,$$

possessing $\dot{\boldsymbol{\gamma}}_\lambda$ as its tangent vector field. The loop $\gamma$ is a continuous counterpart of the so-called *Burgers circuit* considered in crystals with dislocations (e.g. [11]). The curve $\gamma_\lambda$, being an unclosed contour in $E^3$, constitutes then a counterpart of the same circuit in a perfect crystal. The vector defined by this unclosing and running from the finish of the circuit to its start:

(2.17) $$\mathbf{b}[\gamma] = \overrightarrow{\gamma_\lambda(\beta)O} = -\overrightarrow{O\gamma_\lambda(\beta)} = \mathrm{b}^a[\gamma]\mathbf{C}_a, \qquad [\mathbf{b}[\gamma]] = 1,$$

can be called a *Burgers vector* (like that one completing the circuit in a perfect crystal [11]) [13].

If the vector fields $\mathbf{E}_a$ are identified with linear differential operators (Appendix), then the moving coframe $\Phi^* = (\mathrm{E}^a)$ dual to $\Phi$ can be identified with a triple of 1-forms and we have

(2.18) $$\mathrm{b}^a[\gamma] = -\int_\alpha^\beta \dot{\gamma}^a(t)dt = -\oint_\gamma \mathrm{E}^a, \qquad [\mathrm{b}^a[\gamma]] = \mathrm{cm},$$

where Eqs. (2.8) and (2.15)-(2.17) were taken into account. Note that in the paper [13] (and in the papers based on it, e.g. [14], [15] and [16]), the opposite orientation of the Burgers vector has been assumed. The formula

(2.19) $$\mathbf{b}[\gamma] = \mathrm{b}^a[\gamma]\mathbf{C}_a, \qquad \mathrm{b}^a[\gamma] = \varepsilon\oint_\gamma \mathrm{E}^a,$$

where $\varepsilon = \pm 1$, takes into account both these possibilities.

### 3. Material space

Let us consider the base vector fields of a Bravais moving frame $\Phi$ as those that define scales of an *internal length measurement* along local crystallographic directions of the dislocated Bravais crystal (Section 1). For example, the following *intrinsic material metric tensor* $\mathbf{g}$ represents such a length measurement:

(3.1) $$\mathbf{g} = \mathbf{g}[\Phi] = \delta_{ab}\mathrm{E}^a \otimes \mathrm{E}^b = g_{AB}dX^A \otimes dX^B,$$
$$g_{AB} = \delta_{ab}e^a_A e^b_B, \qquad [\mathbf{g}] = \mathrm{cm}^2,$$

where Eqs. (2.3) and (2.5) were taken into account. It is a Riemannian model of the distortion of the globally Euclidean length measurement within a crystalline body $B \subset E^3$ due to many dislocations [2]. Since Riemannian metrics are locally Euclidean, therefore it is an internal length measurement consistent with the observed phenomenon that dislocations have no influence on the local metric properties of the crystalline body (Section 1). Moreover, according to Eqs. (2.12) and (3.1), if $\mathbf{S}[\Phi] = \mathbf{0}$, then the Riemannian space $B_g = (B, \mathbf{g})$ is flat [14]. It means that if dislocations are absent, then the Euclidean internal length measurement induced within the body (Section 1) is not distorted. However, the flatness of the space $B_g$ does not mean a lack of dislocations. For example, the Bravais moving frame $\Phi$ such that

(3.2)
$$[\mathbf{E}_1, \mathbf{E}_2] = \gamma \mathbf{E}_3, \quad [\mathbf{E}_1, \mathbf{E}_3] = -\gamma \mathbf{E}_2, \quad [\mathbf{E}_2, \mathbf{E}_3] = \mathbf{0},$$
$$\gamma = \text{const.} > 0, \quad [\gamma] = \text{cm}^{-1},$$

describes a distribution of dislocations for which the space $B_g$ is flat [14].

Let us consider a *global rescaling* of the internal length measurement defined by:

(3.3)
$$\Phi \mathbf{L} = \left( \mathbf{E}_a L^a{}_b;\ b = 1,2,3 \right), \quad \Phi^* \mathbf{L} = \left( \left( \mathbf{L}^{-1} \right)^a{}_b \mathbf{E}^b;\ a = 1,2,3 \right),$$
$$\mathbf{L} = \left( L^a{}_b;\ {}^{a\downarrow 1,2,3}_{b\to 1,2,3} \right) \in \text{GL}^+(3),$$

where $\text{GL}^+(3)$ denotes the group of all real $3 \times 3$ matrices with positive determinant. It is easy to see that though

(3.4)
$$\mathbf{g}[\Phi \mathbf{L}] = \mathbf{L}^{-1} \mathbf{g}[\Phi] \mathbf{L}^* = g_{ab} \mathbf{E}^a \otimes \mathbf{E}^b,$$
$$g_{ab} = g_{ba} = \text{const.}, \quad \det\left( g_{ab};\ {}^{a\downarrow 1,2,3}_{b\to 1,2,3} \right) > 0, \quad \mathbf{L}^* = \left( \mathbf{L}^{-1} \right)^T,$$

where the constants $g_{ab}$ are parameters independent of $\Phi$, nevertheless the tensorial measure of material anholonomity is invariant under the group $\text{GL}^+(3)$ [8]:

(3.5)
$$\forall \mathbf{L} \in \text{GL}^+(3), \quad \mathbf{S}[\Phi \mathbf{L}] = \mathbf{S}[\Phi].$$

So, without any loss of generality, we can consider the internal length measurement up to its global rescaling. The Riemannian space $B_g = (B, \mathbf{g})$, where $\mathbf{g} = \mathbf{g}[\Phi]$ is defined by Eq. (3.1), will be called a *material space* (associated with the Bravais moving frame $\Phi$). An affinely-invariant Lagrangian description of static self-equilibrium distributions of dislocations based on the condition (3.5) has been proposed in [8].

A Bravais moving frame $\Phi = (\mathbf{E}_a)$ defines, according to Eq. (2.13), the *plastic distortion tensor* $\mathbf{P}$ such that [14]

$$(3.6) \qquad \mathbf{E}_a = \mathbf{P}\mathbf{C}_a, \qquad \mathbf{C}_a \mathbf{c} \mathbf{C}_b = \delta_{ab},$$

where $C = (\mathbf{C}_a;\ a = 1,2,3)$ is a Cartesian basis defined on the Euclidean point space $E^3$ endowed with the Euclidean metric tensor $\mathbf{c}$. The tensor field $\mathbf{P}$ has the form:

$$(3.7) \qquad \mathbf{P} = \mathbf{E}_a \otimes \mathbf{C}^a,$$

where $C^* = (\mathbf{C}^a;\ a = 1,2,3)$ is the Cartesian basis dual to the basis $C$, and possesses the following operational representation (see Appendix):

$$(3.8) \qquad \begin{aligned} \mathbf{P} &= \mathbf{E}_a \otimes \mathbf{C}^a = P^A{}_B \partial_A \otimes dX^B, \qquad P^A{}_B = e^A_{\ a} \overset{a}{C}_B \in C^\infty(B), \\ \mathbf{C}^a &= \overset{a}{C}_B dX^B, \qquad \langle \mathbf{C}^a, \mathbf{u} \rangle = \mathbf{C}^a \mathbf{c} \mathbf{u}, \qquad \mathbf{C}_a \mathbf{c} \mathbf{C}^b = \delta_a^b, \qquad d\mathbf{C}^a = 0, \end{aligned}$$

where the system $\{\mathbf{C}^a;\ a = 1,2,3\}$ of 1-forms constitutes an operational representation of the basis $C^*$ and considered as those defined on the body $B$. Thus, taking into account Eq. (A.12), we obtain the following representation of the Bravais moving co-frame $\Phi^* = (\mathbf{E}^a)$:

$$(3.9) \qquad \mathbf{E}^a = \mathbf{P}^* \mathbf{C}^a, \qquad \mathbf{P}^* = (\mathbf{P}^{-1})^T,$$

and the intrinsic material metric tensor can be written in the form

(3.10)
$$\mathbf{g} = \mathbf{P}^*\mathbf{c}\mathbf{P}^{-1} = \delta_{ab}\mathbf{E}^a \otimes \mathbf{E}^b.$$

Note that in the range of relatively low absolute temperatures $\theta$ (e.g. $\theta \leq 0,6\theta_m$, $\theta_m$ - melting point), the *plastic deformation* of metals is primarily caused by *gliding* on (local) crystallographic slip systems. This type of deformation is inevitably connected with (local) *rotations* of crystallographic directions (cf. the Kondo Gedanken Experiment - Section 2). The deformation (of a Bravais monocrystal) can be composed of the gliding itself (see Section 1), corresponding to the *plastic deformation* tensor $\mathbf{F}_p$ does not changing of the local crystallographic directions, that is

(3.11)
$$\mathbf{F}_p \mathbf{E}_a = \mu_a \mathbf{E}_a, \quad \mu_a > 0, \quad a = 1,2,3,$$

and the *elastic deformation* $\mathbf{B}$ changing these directions (cf. Eq. (2.7)):

(3.12)
$$\mathbf{e}_a = \mathbf{B}\mathbf{E}_a.$$

The *total deformation tensor* $\mathbf{F}$ (Section 2) can be represented now in the form:

(3.13)
$$\mathbf{F} = \mathbf{B}\mathbf{F}_p.$$

Particularly, if $\xi = (\xi^a)$, $[\xi^a] = cm$, is a curvilinear coordinate system on $B$ and

(3.14)
$$\mathbf{F}_p = \mathbf{P} = \mathbf{E}_a \otimes \mathbf{c}^a,$$
$$\mathbf{c}^a \mathbf{c}\mathbf{c}_b = \delta^a_b, \quad \partial/\partial \xi^a = \partial_{\mathbf{c}_a},$$

then

(3.15)
$$\mathbf{E}_a = \mathbf{P}\mathbf{c}_a, \quad a = 1,2,3,$$

and Eq. (3.11) is fulfilled iff there are positive scalars $\mu_a \in C^\infty(B)$ such that

(3.16)
$$\mathbf{E}_a = \mu_a \mathbf{c}_a, \quad a = 1,2,3.$$

Eq. (3.13) takes then the form

(3.17)
$$\mathbf{F} = \mathbf{BP},$$
$$\mathbf{P} = P^b{}_a \mathbf{c}_b \otimes \mathbf{c}^a, \quad P^b{}_a = \mu_a \delta^b_a.$$

For example, if there exists a coordinate system $\xi = (\xi^a)$ on $B$ such that

(3.18)
$$E^a = e^{-\sigma} d\xi^a, \quad a = 1, 2, 3,$$
$$\langle d\xi^a, \partial_b \rangle = \delta^a_b, \quad \partial_a = \partial/\partial \xi^a = \partial_{c_a},$$

what means that the material space $B_g$ is a *conformally flat* Riemannian manifold

(3.19)
$$\mathbf{g} = e^{-2\sigma} \mathbf{c}, \quad \mathbf{c} = \delta_{ab} d\xi^a \otimes d\xi^b,$$

then it is the case of Eq. (3.16) with

(3.20)
$$\mu_a = e^{\sigma}, \quad a = 1, 2, 3.$$

The case of *secondary vacancies* or *secondary interstitials* can be modeled then by the condition $\sigma > 0$ or $\sigma < 0$, respectively; if $\sigma = 0$, then dislocations are absent [6].

More generally, a transformation $\mathbf{g} \to \mathbf{g}_\alpha$ of the metric tensor $\mathbf{g}$ defined by

(3.21)
$$\mathbf{g}_\alpha = \alpha \mathbf{g},$$
$$\alpha = e^{-2\sigma}, \quad \alpha \in C^\infty(B),$$

is called *conformal*. If additionally there exists a vector field $\mathbf{u} \in W(B)$ such that

(3.22)
$$\mathbf{g}_\alpha = \boldsymbol{\varepsilon},$$

where $\boldsymbol{\varepsilon}$ is an infinitesimal strain (see Appendix) then the transformation is called *infinitesimally conformal*. An infinitesimally conformal transformation such that

(3.23)
$$\varepsilon_{AB} = \nabla^g_A u_B = \nabla^g_B u_A,$$

where $\nabla^g$ denotes the Levi-Civita covariant derivative on $B_g$ (see Appendix), defines a particular case of the so-called *equidistant material space* (Section 6).

## 4. Dislocation densities

It seems physically reasonable to take into account the influence of secondary point defects on the Burgers vector. For example, the use of internal length measurement scales (Sections 1 and 3) for computing its components can be helpful here. Namely, let us consider a Burgers circuit $\gamma \subset B$ (Section 2) as the one located in the *Riemannian material space* $B_g = (B, \mathbf{g})$, where $\mathbf{g} = \mathbf{g}[\Phi]$ is the intrinsic material metric tensor associated with a Bravais moving frame $\Phi = (\mathbf{E}_a)$ (Section 3). Next, let us identify the base vector fields $\mathbf{E}_a$ with linear differential operators and let $\Phi^* = (E^a)$ denote the triple of base 1-forms dual to the such understood $\Phi$ (Appendix). Then, the integrals of Eq. (2.19) that define components $b^a[\gamma]$ of a Burgers vector $\mathbf{b}[\gamma]$, can be treated as mappings $\gamma \subset B_g \to \mathbf{b}[\gamma] \in E^3$ defined on the space $B_g$ (e.g. [17]). Let $\Sigma \subset B$ be a surface possessing a closed contour $\gamma$ as its boundary and treated as a two-dimensional compact, connected and oriented Riemannian submanifold of $B_g$ (that is endowed with a Riemannian metric induced from $B_g$). Since

$$\text{(4.1)} \qquad b^a[\gamma] = \varepsilon \int_\Sigma \tau^a,$$

where Eqs. (2.4) and (2.19) were taken into account and, according to Eqs. (2.10) and (2.11), we have

$$\text{(4.2)} \qquad \mathbf{S}[\Phi] = \tau^a \otimes \mathbf{E}_a, \quad \tau^a = dE^a = S_{bc}{}^a \Omega^{bc}, \quad S_{bc}{}^a = -\frac{1}{2} C_{bc}^a,$$

$$\Omega^{bc} = \frac{1}{2}\left(E^b \otimes E^c - E^c \otimes E^b\right), \quad [\Omega^{bc}] = cm^2, \quad [S_{bc}{}^a] = cm^{-1},$$

the Stokes theorem states that [17]:

(4.3) $$\int_\Sigma \tau^a = \int_\Sigma S_{bc}{}^a d\Sigma^{bc}, \qquad d\Sigma^{bc} = l^{bc} d\Sigma, \qquad l^{bc} = e^{bcd} l_d, \qquad l_d = \delta_{de} l^e,$$

where $d\Sigma$, $[d\Sigma] = cm^2$, denotes the surface element of $\Sigma$ normal to the unit vector

$\mathbf{l} = l^a \mathbf{E}_a = l^A \partial_A \in W(B)$ tangent to $B_g$ (i.e. $\|\mathbf{l}\|_g = l^a l_a = 1$, $[l^a] = 1$) and

(4.4) $$e^{abc} = e_A^{\ a} e_B^{\ b} e_C^{\ c} e^{ABC}, \qquad e^{ABC} = g^{-1/2} \varepsilon^{ABC},$$
$$g = \det\left(g_{AB};\ {}^{A\downarrow 1,2,3}_{B\to 1,2,3}\right) = e^{-2}, \qquad e = e_\Phi = \det\left(e^{\ A}_a;\ {}^{A\to 1,2,3}_{a\downarrow 1,2,3}\right),$$

where $\varepsilon^{ABC}$ denotes the permutation symbol associated with the coordinate system $X = (X^A)$ and $e^{ABC}$ is a contravariant 3-vector density of weight +1; $e^{abc} \doteq \varepsilon^{abc}$ denote the permutation symbols associated with the Bravais moving frame $\Phi = (\mathbf{E}_a)$ and considered as components in this base of a contravariant 3-vector density of weight +1 in $B_g$. It follows from Eqs. (4.1)-(4.4) that [12]:

(4.5) $$b^a[\gamma] = \int_\Sigma \alpha^{ba} d\Sigma_b, \qquad d\Sigma_b = d\Sigma l_b,$$

where

(4.6) $$\alpha^{ba} = \varepsilon S_{cd}{}^a e^{cdb}, \qquad [\alpha^{ba}] = cm^{-1}.$$

The tensor

(4.7) $$\boldsymbol{\alpha} = \alpha^{ab} \mathbf{E}_a \otimes \mathbf{E}_b, \qquad [\boldsymbol{\alpha}] = cm^{-3}$$

is, up to the choice of the Burgers vector orientation, a Riemannian modification of the so-called *dislocation density tensor* (or the *Ney's tensor* - [18]) considered in the literature. Likewise, the *scalar volume dislocation density* $\rho$ of a finite total length $L_d(B)$ of dislocation lines located in $B$ will be measured with respect to the material volume element:

(4.8)
$$0 < L_d(B) = \int_B \rho \omega_g = \int_B \rho dV_g,$$
$$\omega_g = E^1 \wedge E^2 \wedge E^3 = e dX^1 \wedge dX^2 \wedge dX^3, \qquad dV_g = \sqrt{g} dX^1 dX^2 dX^3,$$

where $[\rho] = \text{cm}^{-2}$, $[L_d(B)] = \text{cm}$, $\omega_g$ is the volume 3-form of $B_g$ and $dV_g$ denotes the material volume element. A distribution of dislocations for which $\alpha^{ab} = \text{const.}$ and $\rho = \text{const.}$ is called *uniformly dense* [15]. In this case all possible types of distortions of lattice lines can be described by the well-known Bianchi classification of three-dimensional real Lie algebras (see e.g. [19]). Consequently, we can distinguish *Bian-chi-type distortions* of continuized Bravais crystals produced by distributions of dislocations [15]. These distortions can be interpreted as *fundamental states* of these de-fective crystalline solids. [8].

Let us write the components $\alpha^{ab}$ of the dislocation density tensor in the form:

(4.9)
$$\alpha^{ab} = \gamma^{ab} + \sigma^{ab},$$
$$\gamma^{ab} = \alpha^{(ab)}, \qquad \sigma^{ab} = \alpha^{[ab]} = \frac{1}{2} t_c e^{cab},$$

where, according to Eqs. (4.2) and (4.6), we have:

(4.10)
$$t_a = e_{abc} \alpha^{bc} = \varepsilon C_{ab}^b, \qquad \varepsilon = \pm 1,$$

and $e_{abc} \doteq \varepsilon_{abc} (= \varepsilon^{abc})$ denote the permutation symbols considered as components in the base $\Phi^* = (E^a)$ of a covariant 3-vector density of weight $-1$ in $B_g$. It follows from Eqs. (4.2), (4.6), (4.9), and (4.10) that

(4.11)
$$\varepsilon C_{ab}^c = t_{[a} \delta_{b]}^c - e_{abd} \gamma^{dc}.$$

Therefore, the long-range distortion of a continuously dislocated Bravais crystal with secondary point defects describes the following pair $(\gamma, \mathbf{t})$ of objects defined on $B_g$:

(4.12)
$$\gamma = \gamma^{ab} \mathbf{E}_a \otimes \mathbf{E}_b, \qquad \gamma^{ab} = \gamma^{ba},$$
$$\mathbf{t} = t^a \mathbf{E}_a, \qquad t^a = \delta^{ab} t_b; \qquad [\gamma^{ab}] = [t^a] = \text{cm}^{-1}.$$

## 5. Self-balance equations

Let $\nabla^g$ denote the Levi-Civita covariant derivative (see Appendix) in the Riemannian material space $B_g$ (Section 3). If the Bravais moving frame $\Phi = (\mathbf{E}_a)$ defined by Eq. (2.5) is considered as a system of vector fields tangent to $B_g$, then [20]

$$\nabla^g \mathbf{E}_a = \nabla^g_B \underset{a}{e^C} dX^B \otimes \partial_C = \omega^b{}_a \otimes \mathbf{E}_b, \qquad \omega^b{}_a = \omega_c{}^b{}_a E^c,$$
(5.1)
$$\omega_b{}^a{}_c - \omega_c{}^a{}_b = C^a_{bc}, \qquad \omega_b{}^a{}_c = \lambda^d{}_b \varepsilon_d{}^a{}_c, \qquad \varepsilon_d{}^a{}_c = \delta^{ab}\varepsilon_{dbc},$$

where $\lambda^d{}_b \in C^\infty(B)$. Therefore

$$\nabla^g_B \underset{a}{e^C} = \underset{b}{e_B} \underset{c}{e^C} \omega_b{}^a{}_c,$$
(5.2)
$$\omega_b{}^a{}_a = 0, \qquad \omega_a{}^a{}_b = -C^b_{ba}.$$

It follows from the definition of the divergence operator in $B_g$ that

(5.3) $$\mathrm{div}_g \mathbf{E}_a = \nabla^g_A \underset{a}{e^A} = g^{-1/2} \partial_A \left( g^{1/2} \underset{a}{e^A} \right), \qquad g = \det(g_{AB}).$$

Thus, the following *self-balance equations* hold [15]:

(5.4) $$\mathrm{div}_g \mathbf{E}_a = -\sigma_a, \qquad \sigma_a = \varepsilon t_a = C^b_{ab}, \qquad a = 1, 2, 3.$$

Let us return to the definition of the Burgers vector $\mathbf{b}[\gamma]$ given by Eqs. (2.4), (2.19), (4.1), (4.5), and (4.6). If $U \subset B \subset E^3$ is a three-dimensional regular region with a regular closed boundary $\Sigma$, then the vector field $\mathbf{F}[\Sigma]$ defined by [13]

(5.5) $$\mathbf{F}[\Sigma] = F^a[\Sigma] \mathbf{C}_a, \qquad F^a[\Sigma] = \varepsilon \int_\Sigma \tau^a,$$

is a modification of the so-called *Frank vector*. Rewriting Eq. (4.7) in the form

(5.6) $$\boldsymbol{\alpha} = \boldsymbol{\alpha}^a \otimes \mathbf{E}_a, \qquad \boldsymbol{\alpha}^a = \alpha^{ba} \mathbf{E}_b = \alpha^{Ba} \partial_B$$

and taking into account the divergence theorem of Gauss [3], we obtain that

$$\text{(5.7)} \qquad F^a[\Sigma] = \int_\Sigma (\boldsymbol{\alpha}^a, \mathbf{l})_g \, d\Sigma = \int_U \text{div}_g \boldsymbol{\alpha}^a \, dV_g,$$

where $dV_g$ denotes the Riemannian volume element of $B_g$, $d\Sigma$ denotes the surface element of the surface $\Sigma \subset B_g$, $\mathbf{l}$ is the unit outer normal vector field on $\Sigma$, and

$$\text{(5.8)} \qquad \text{div}_g \boldsymbol{\alpha}^a = \nabla_B^g \alpha^{Ba} = g^{-1/2} \partial_B \left( g^{1/2} \alpha^{Ba} \right).$$

Since $d\tau^a = 0$, we obtain that the following *self-balance equations* hold [13]:

$$\text{(5.9)} \qquad \text{div}_g \boldsymbol{\alpha}^a = 0, \qquad a = 1, 2, 3,$$

i.e., the modified Frank vector vanishes. Note, that if $\mathbf{g}$ is a flat metric, then $B_g$ can be treated (at least locally) as a submanifold of the Euclidean point space $E^3$. Eq. (5.9) is then usually interpreted as the one stating that lines of dislocations do not terminate within the crystal [21]. It means that the dislocations under consideration must either form closed loops or branch into other dislocations [11]. Therefore, Eq. (5.9) can be interpreted as the one admitting the appearance of dislocation lines terminating within the crystal due to the occurrence of secondary point defects in this crystal [13].

## 6. Crystal surfaces

Let $B_g = (B, \mathbf{g})$ be the material space associated with a Bravais moving frame $\Phi = (\mathbf{E}_a)$ and let us consider a two-dimensional distribution $\pi = \{\pi_p, p \in M\}$ on $B_g$ (Appendix) of *local crystal planes* (Section 1). A two-dimensional distribu-tion (of planes) is called *integrable* if there exists a family $\Pi = \{\Sigma_p, p \in M\}$ of two-dimensional submanifolds of $B_g$, called *integral manifolds* of $\pi$, such that $p \in \Sigma_p$ and for each $q \in \Sigma_p$ the plane $\pi_q$ is tangent to $\Sigma_p$ at $q$ (Appendix). These integral manifolds can be considered as *crystal surfaces* of a continuized defective Bravais crystal $B \subset E^3$ (Sections 1 and 2) isometrically embedded (at least locally) in the Riemannian material space $B_g$ [22]. Let $\mathbf{E}_\alpha(p) \in \pi_p$, $\alpha = 1,2$, be a base of the vector space $\pi_p \in \pi$. The distribution $\pi$ is *involutive* if there are smooth scalars $C_{\alpha\beta}^\kappa$, $\alpha, \beta, \kappa = 1,2$, on $B_g$ such that Eq. (A.13) holds (see Appendix, Theorem). A distribution is involutive iff it is *integrable* or, eqivalently [17], [23], the system of equations

(6.1) $$\mathbf{E}_\alpha \varphi = \mathrm{e}_\alpha^A \partial_A \varphi = 0, \quad A = 1,2,3; \; \alpha = 1,2,$$

where Eq. (2.5) was taken into account, has a solution that defines surfaces of the family $\Pi$ of *integral manifolds* as those given by:

(6.2) $$\Sigma_c = \varphi^{-1}(c), \quad d\varphi \neq 0,$$

where $c \in \mathbb{R}$ is a constant, i.e., $\Pi = \{\Sigma_c, c \in \mathbb{R}\}$. It can be shown that for each $p \in B$ there are then a coordinate neighborhood $U$ of $p$ and a coordinate system $X = (X^A)$ on $U$ such that $X^3 = \varphi$. For any such coordinates, the ordered set of vector fields $\{\partial_\alpha = \partial/\partial X^\alpha, \; \alpha = 1,2\}$ is a local basis for $\Pi$ and the slices

(6.3) $$\Sigma_c = \{q \in U : \; X^3(q) = c\},$$

belong to $\Pi$ [17]. Consequently, we can consider, at least locally, an involutive distribution of local crystal planes as the one in which integral manifolds $\Sigma_c$ are defined by a *coordinate system* on $B$. Moreover, it is known that $B_g$ is *foliated* by the distribution $\pi$, that is, through each point $p \in B_g$ there passes an unique maximal integral manifold of $\pi$ [17].

Thus, let $(B_g, \varphi)$ be a system consisting of the material Riemannian space $B_g$ and a scalar $\varphi \in C^\infty(B)$ of Eq. (6.2). Since $\varphi \neq \text{const.}$, we can choose a coordinate system $(U, X)$, $X = (X^C)$, such that for $X = X(p) \in \mathbb{R}^3$, $p = p(X) \in U$, we have:

$$(6.4) \quad \begin{aligned} \mathbf{g}(p(X)) &= g_{AB}(X^C(p)) dX^A_{p(X)} \otimes dX^B_{p(X)} \\ &\doteq g_{\alpha\beta}(X^C) dX^\alpha \otimes dX^\beta + g_{33}(X^C) dX^3 \otimes dX^3 \equiv \mathbf{g}(X), \end{aligned}$$

where $\alpha, \beta = 1, 2$, $\doteq$ means that a relation is defined using a distinguished coordinate system, it is designated $dX^A \equiv dX^A_{p(X)}$, $A = 1, 2, 3$, and

$$(6.5) \quad \varphi(p(X)) \doteq \varphi(X^3), \qquad g_{AB}\varphi^A\varphi^B \neq 0, \qquad \varphi_A = \partial_A \varphi = g_{AB}\varphi^B,$$

where it was identified $\varphi|U$ with $\varphi \circ X^{-1} : X(U) \to \mathbb{R}$. Since, according to Eq. (6.5), the function $\varphi|U$ has an inverse differentiable mapping $(\varphi|U)^{-1}$, the slices $\Sigma_c \cap U$ can be assumed, without any loss of generality, to be coordinate surfaces of Eq. (6.3) for a coordinate system $X = (X^C; C=1,2,3)$ such that $X^3 = \varphi|U$. Moreover, we will assume that the intrinsic material metric tensor $\mathbf{g}$ of Eq. (6.4) can be reduced by a suitable transformation of coordinates to the so-called *geodesic form* [24]:

$$(6.6) \quad \mathbf{g}(X) \doteq g_{\alpha\beta}(X^C) dX^\alpha \otimes dX^\beta + dX^3 \otimes dX^3.$$

The surfaces $\Sigma_c$, $c \in \mathbb{R}$ are said then to be *geodesically parallel* to the surface $\Sigma_0$.

The existence of the above-mentioned transformation of coordinates can be related to the existence of point transformations in $B_g$ leaving **g** as well as $\varphi$ invariant. Such a transformation is called a *scalar-preserving isometry* $B_g$ or, more concretely, of the system $(B_g, \varphi)$ [25]. Namely, let us denote $\mathbf{u} = u^A \partial_A$ an *infinitesimal scalar-preserving isometry*. It follows from the definition that the following equations must hold good:

(6.7) $\qquad 2\varepsilon_{AB} = \nabla^g_A u_B + \nabla^g_B u_A = 0, \qquad u^A \varphi_A = 0, \qquad u_A = g_{AB} u^B,$

where $\varepsilon_{AB}$ are components of the infinitesimal strain tensor (see Appendix). Next, let us observe that the hypersurface $\Sigma_c$ of Eq. (6.2) can be regarded as a 2-dimensional Riemannian space endowed with the metric tensor $\mathbf{a}_c$ induced from $B_g$, that is, in the coordinate system of Eq. (6.4) we have:

(6.8) $\qquad \mathbf{a}_c(X^\kappa) = a_{c,\alpha\beta}(X^\kappa) dX^\alpha \otimes dX^\beta, \qquad a_{c,\alpha\beta}(X^\kappa) \doteq g_{\alpha\beta}(X^\kappa, c),$

where $\alpha, \beta, \kappa = 1, 2$. The Riemannian manifolds $(\Sigma_c, \mathbf{a}_c)$ in general are not isometric. The conditions of Eq. (6.7) can be expressed now in the following form:

(6.9) $\qquad \nabla^c_\alpha u_\beta + \nabla^c_\beta u_\alpha = 0, \qquad \partial_3 u^\kappa = 0, \qquad u^\kappa \partial_\kappa g_{33} = 0, \qquad u^3 = 0,$

where $\nabla^c$ denotes the Levi-Civita covariant derivative with respect to the metric tensor $\mathbf{a}_c$ and $u_\alpha = a_{c,\alpha\beta} u^\beta$, so that $u_\alpha = u_\alpha(X^\kappa, c)$ even if $u^\alpha = u^\alpha(X^\kappa)$. A group is said to be *transitive* when, by means of its transformations, any ordinary point of $B_g$ can be transformed into any other ordinary point; otherwise it is *intransitive*. For example, if $\Sigma_c$ admits a group of isometries of order 2, then the group is transitive. If a scalar-preserving isometry group is transitive on each hypersurface $\Sigma_c$, then $g_{33}$ of Eq. (6.5) can be reduced to one by a suitable transformation of coordinates [25].

Let the unit vector field $\mathbf{n} = n^A \partial_A$ in $B_g$ be normal to the surfaces $\varphi = \text{const.}$. It follows from Eq. (6.2) that then we are dealing with crystal surfaces such that

(6.10) $$\mathbf{n} = \mathbf{E}_3.$$

Thus, in the coordinate system of Eq. (6.6), we have

(6.11) $$\mathbf{E}_3 \doteq \partial_3, \qquad n^A \doteq \delta_3^A,$$

and surfaces $\Sigma_c$, $c \in \mathbb{R}$, can be characterized as made up of endpoints of geodesics of the same length tangent to $\mathbf{n}$ and starting e.g. from the surface $\Sigma_0$. That is why further on, the material space $B_g$ endowed with the intrinsic metric tensor $\mathbf{g}$ admitting its representation in the geodesic form is called an *equidistant material space* (a par-ticular case of such a material space has been considered in [14], [16] and [26]). If additionally the following condition is fulfilled:

(6.12) $$g_{\alpha\beta}(X^A) \doteq \Psi(X^A) a_{\alpha\beta}(X^\kappa),$$

then, according to Eq. (6.8), the components $a_{c,\alpha\beta}$ and $b_{c,\alpha\beta}$ of the first and second fundamental forms of $\Sigma_c$ are given respectively by:

(6.13) $$a_{c,\alpha\beta}(X^\kappa) = \Psi(X^\kappa, c) a_{\alpha\beta}(X^\kappa),$$

and [24]

(6.14) $$b_{c,\alpha\beta}(X^\kappa) = H(X^\kappa, c) a_{c,\alpha\beta}(X^\kappa), \qquad H = -\frac{1}{2}\partial_3 \Psi,$$

where $H_c(X^\kappa) = H(X^\kappa, c)$ is the *mean curvature* of the surface $\Sigma_c$ embedded in $B_g$ and the definition of the mean curvature according to SCHOUTEN [9], in place of the definition of EISENHART [24] was taken into account. Moreover, the crystal surfaces $\Sigma_c$ are then *umbilical*, that is these are a generalization of a plane (or sphere) in the Euclidean 3-space $E^3$. For example, it is the case of a Bravais moving frame such that

$$(6.15) \quad \mathbf{E}_\alpha(p(X)) \equiv \mathbf{E}_\alpha(X^C(p)) \doteq \Psi^{-1/2}(X^3(p))\mathbf{a}_\alpha(X^\kappa(p)), \qquad \mathbf{E}_3 \doteq \partial_3,$$

where the vector fields $\mathbf{E}_a$ defined on $U \subset B$ are identified with the vector fields $\mathbf{E}_a \circ X^{-1}$ defined on $X(U) \subset \mathbb{R}^3$, and it was denoted

$$(6.16) \quad \Psi(X^3) = a^2 e^{-2h(X^3)}, \qquad \Psi(0) = 1, \qquad a = \text{const.}$$

The metric tensors of the crystal surfaces $\Sigma_{\mathbf{a}_c} = (\Sigma_c, \mathbf{a}_c)$ are given then by

$$(6.17) \quad \mathbf{a}_c = \Psi(c)\mathbf{a}, \qquad \mathbf{a} = \mathbf{a}(X^\kappa) = a_{\alpha\beta}(X^\kappa)dX^\alpha \otimes dX^\beta,$$

and their mean curvatures have the form

$$(6.18) \quad H_c = H(c) = h'(c),$$

where $h' = dh/dX^3$. These crystal surfaces are umbilical with the *constant mean curvature* $H_c$. Moreover, since in the coordinate description of Eq. (6.6) we have:

$$(6.19) \quad \mathbf{g}(X) \doteq \Psi(X^3)\mathbf{a}(X^\kappa) + dX^3 \otimes dX^3,$$

it is, for example, the case of *infinitesimally conformal equidistant material spaces* (Section 3) considered in [14], [16] and [26].

If, in the case of Eq. (6.19), the scalar-preserving isometry group has its maximal order (which equals here 3 – [25]), then any crystal surface $\Sigma_c$, $c \in \mathbb{R}$, has additionally the *constant Gaussian curvature* $K_c$ [25] and thus we have [24]

$$(6.20) \quad \begin{aligned} \mathbf{a}_c(X^\kappa) &\doteq \left(1 + \frac{1}{4}K_c r^2\right)^{-2} \mathbf{h}(X^\kappa), \qquad \mathbf{h}(X^\kappa) = \delta_{\alpha\beta}dX^\alpha \otimes dX^\beta, \\ r^2 &= \delta_{\alpha\beta}X^\alpha X^\beta, \qquad \alpha, \beta = 1, 2; \qquad K_c = \Psi(c)^{-1}K_0, \end{aligned}$$

where Eq. (6.19) was taken into account, and the Killing vectors has the following form [25]:

$$(6.21) \quad \mathbf{u}_\kappa = \underset{\kappa}{u}^\alpha \partial_\alpha, \quad \underset{\kappa}{u}^\alpha \doteq \left(1 - \frac{1}{4} K_c r^2\right) \delta_\kappa^\alpha + \frac{1}{2} K_c X_\kappa X^\alpha, \quad \mathbf{u}_3 \doteq X^2 \partial_1 - X^1 \partial_2,$$

$$r^2 = \delta_{\alpha\beta} X^\alpha X^\beta, \quad X_\kappa = \delta_{\kappa\lambda} X^\lambda, \quad \alpha, \lambda, \kappa = 1, 2.$$

Eq. (6.21) has the following physical interpretation [14], [26]. Let a crystal surface $\Sigma_c$ be a *glide surface* (Section 1). If it is a *parabolic* surface ($K_c = 0$), then $\Sigma_c$ admits as its motions, in the small at least, the deformations of Euclidean plane characterizing the *single glide* case (Section 1): planar rotations and translations [4]. In the hyperbolic case ($K_c < 0$), the motions of $\Sigma_c$ constitute the three-dimensional Lorentz group. It is easy to see that the particular Lorentz transformations can be considered as a deformation of Euclidean plane changing a square into a rhomb (it is the so-called *pure shear*) [27]. The remaining three-dimensional Lorentz transformations are planar Euclidean rotations or their compositions with pure shearing. The case of elliptic *glide surface*s ($K_c > 0$) can be considered as the one corresponding to an elementary act of plasticity connected with the phenomenon of crystal fragmentation in the plastic yielding process and called *rotational plasticity* [28].

Note that, in contrast with the traditional approach in which the crystal surfaces of defective crystals are considered to be located in the Euclidean ambient space of the body, here these surfaces are considered as submanifolds of a material Riemannian space. Consequently, such crystal surfaces can be only locally isometrically embedded into this Euclidean space. However, although the Gaussian curvature is preserved under this embedding, the mean curvature is not preserved. Particularly, if the material space has a scalar constant curvature $K_g$, then [24]

$$(6.22) \qquad K_c = H_c^2 + K_g.$$

It suggests that the mean curvature of crystal surfaces embedded in a material space has the physical meaning of a *material parameter* being a measure of the influence of secondary point defects on the glide phenomenon.

## 7. Final remarks

It is known that the work done in plastic deformations, more than 90% goes into heat and less than 10% into energy stored for instance in a densification of dislocations (nevertheless, the 10% are of great importance since they are responsible for the work-hardening of the body) [7]. It suggests to consider a metrical relationship in the material space (Section 3) determined by specifying the change in the internal length measurement of a vector due to the phenomenon of thermal extension. Particularly, if the body is *thermally isotropic*, then the thermal extension is the same in all directions at every point $p \in B$ and the change in the measure of length of a vector can be described e.g. in the framework of the so-called Weyl geometry. Namely, let us denote, for $\mathbf{v} \in W(B)$ and $p \in B$, the length of the vector $\mathbf{v}_p = \mathbf{v}(p) \in T_p(B)$ by

$$(7.1) \qquad \begin{aligned} l_g(\mathbf{v}_p) &= l_g(\mathbf{v})(p) = l_{g,p}(\mathbf{v}), \\ l_{g,p(X)}^2(\mathbf{v}) &= g_{AB}(X)v^A(X)v^B(X), \end{aligned}$$

where $p = p(X)$ iff $X = X(p) \in \mathbb{R}^3$ and $\mathbf{v}_{p(X)} = v^A(X)\partial_{A|p(X)}$ in a coordinate system $(U, X)$, $p \in U$, on $B$. Next, let us consider the infinitesimal variation operator $\delta$

defined by Eq. (A.22) where $\nabla$ is a symmetric covariant derivative (see e.g. [20]). The Weyl geometry is defined by the condition [9]:

(7.2) $$\delta g_{AB} = g_{AB}\kappa, \quad \kappa = \kappa_A dX^A, \quad \kappa_A \in C^\infty(B).$$

Since

(7.3) $$\delta l_g^2(\mathbf{v}) = 2l_g(\mathbf{v})dl_g(\mathbf{v}) \\ = 2v_A \delta v^A + \delta g_{AB} v^A v^B, \quad v_A = g_{AB} v^B,$$

we obtain, according to Eqs. (7.1)-(7.3), that in the Weyl geometry, if $\mathbf{v} \in W(B)$ is a *covariant constant* defined by Eq. (A.23), then

(7.4) $$dl_g(\mathbf{v})/l_g(\mathbf{v}) = \varepsilon, \quad \varepsilon = \kappa/2,$$

independently of the choice of a $\nabla$- covariantly constant vector field $\mathbf{v}$. Consequently, we can take e.g.

(7.5) $$\varepsilon = \varepsilon(\theta, d\theta) = \varepsilon_A(\theta, d\theta) dX^A,$$

where $\theta \in C^\infty(B)$, $\theta \geq 0$, is a field of absolute temperatures of the body $B$, as a *constitutive relation* defining the thermal distortion of the internal length measurement associated with a plastic deformation in the thermally isotropic defective Bravais crystal. Note that if in $(B_g, \nabla)$ the intrinsic material metric tensor $\mathbf{g}$ undergoes a *conformal transformation* of Eq. (3.21), then

(7.6) $$\delta(\alpha g_{AB}) = \alpha g_{AB} \kappa_\alpha, \quad \kappa_\alpha = \kappa + d\ln\alpha.$$

Hence if we take $\mathbf{g}_\alpha$ as the metric tensor instead of $\mathbf{g}$, and if at the same time $\kappa$ is transformed into $\kappa_\alpha$ we get the same covariant derivative [9]. In particular, if

(7.7) $$\varepsilon_A(\theta, d\theta) = \beta(\theta)\partial_A \theta,$$

where $\beta(\theta)$ is the coefficient of linear thermal expansion, then

(7.8) $$\kappa(\theta, d\theta) = 2\varepsilon(\theta, d\theta) = d\lambda(\theta), \qquad \lambda(\theta) = 2\int_{\theta_0}^{\theta} \beta(\vartheta) d\vartheta,$$

and for

(7.9) $$\alpha = e^{-2\sigma}, \qquad \sigma = \lambda/2,$$

we obtain that

(7.10) $$\nabla(\alpha g_{AB}) = 0,$$

that is, in this particular case, the thermal distortion of the internal length measurement takes the form of a conformal transformation of the intrinsic material metric tensor **g** (see Section 3, Example). Then $\nabla = \nabla^{g_\alpha}$ is the Levi-Civita covariant derivative associated with the metric tensor $\mathbf{g}_\alpha$ and the Bravais moving frame $\Phi = (\mathbf{E}_a)$ is transformed into $\Phi_\theta = (\mu(\theta)\mathbf{E}_a)$ where

(7.11) $$\mu(\theta) = e^{\sigma(\theta)}.$$

For example, since we are considering an internal length measurement, it seems reason-able to introduce a *characteristic length* $l(\theta)$ of the thermally isotropic defective Bravais crystal by the rule (cf. [29]):

(7.12) $$\beta(\theta) = \frac{1}{l(\theta)} \frac{dl(\theta)}{d\theta}.$$

Then $\mu(\theta) = l(\theta)/l(\theta_0)$.

In the thermodynamic theory of plasticity are introduced the so-called *internal variables* in order to describe various internal states of the material in a unified framework (e.g [30]). For example, a general (but physically indefinite) Riemannian metric is considered in the paper [31] as such a variable. The above discussed pair $(\mathbf{g}, \varepsilon)$ constrained by Eqs. (7.5) and (7.6) can be also taken as an internal variable.

**Appendix**

Let $B \subset E^3$ denote a *body* identified with its distinguished spatial configuration being an open and contractible to a point subset of the three-dimensional Euclidean point space $E^3$ [2]. We will consider coordinate systems $X = (X^A)$ defined on open subsets $U \subset B$ and such that $[X^A] = \text{cm}$. Let $\sigma_A$ denote the following curve:

(A.1)
$$\sigma_A : (-\varepsilon,\ \varepsilon) \ni s \to X^{-1}(X(p) + s\boldsymbol{\delta}_A) \in B,$$
$$\boldsymbol{\delta}_A = (\delta_{AB};\ B \to 1,2,3), \quad \sigma_A(0) = p, \quad [s] = \text{cm},$$

with the tangent vector $\mathbf{c}_A(p)$ defined by:

(A.2)
$$\mathbf{c}_A(p) = \dot{\sigma}_A(0) \in \boldsymbol{E}^3, \quad [\mathbf{c}_A] = \text{cm}^{-1},$$

where $\dot{\sigma}_A = d\sigma_A / ds$ and $\boldsymbol{E}^3$ denotes a three-dimensional Euclidean vector space identified with the space of all translations in $E^3$. For a coordinate mapping $(U, X)$ and for a point $p \in U$, the curves of Eq. (A.1) are called *coordinate curves* passing through $p$. The fields $\partial_{\mathbf{c}_A} \equiv \partial_A : U \ni p \to \partial_{A|p}$ of differential operators, where

(A.3)
$$\forall f \in C^\infty(U), \quad \partial_{\mathbf{c}_A(p)}(f) \equiv \partial_{A|p}(f) = \frac{\partial(f \circ X^{-1})}{\partial X^A}(X(p)),$$

are called *associated* with the chart $(U, X)$. The operator $\partial_{A|p}$ defines an *operational representation* of the vector $\mathbf{c}_A(p)$ tangent to the coordinate curve $\sigma_A$ passing through the point $p$. More generally, an arbitrary chosen vector

(A.4)
$$\mathbf{v}_p = v^A(p)\mathbf{c}_A(p) \in \boldsymbol{E}^3,$$

tangent to *B* at *p*, can be identified with the operator $\partial_{v_p}$ of the differentiation in the direction $v_p$ defined by the following rule:

$$(A.5) \qquad \forall f \in C^\infty(U), \qquad \partial_{v_p} f = \frac{\partial (f \circ X^{-1})}{\partial X^A}(X(p)) v^A(p).$$

Thus, the vector fields $c_A$ can be identified with their operational representations $\partial_A$ and a vector field $\mathbf{v} = v^A c_A$ can be identified with the operator $\partial_v$ defined by

$$(A.6) \qquad \forall f \in C^\infty(B), \forall p \in B, \qquad \partial_v f(p) = \partial_{v_p} f.$$

Further on, in order to simplify the notations, we will denote this differential linear op-erator as $\mathbf{v} = v^A \partial_A$, $v^A \in C^\infty(B)$. Such a vector field is called *contravariant*. The fields $v = v_A dX^A$ of 1-forms on *B* are called *covariant* vector fields.

Let W(*B*) denote the set of all smooth vector fields on *B* tangent to *B* and identified with the linear first-order differential operators. Since W(*B*) is closed under addition of these vector fields and their multiplication by real numbers and by smooth functions defined on *B*, it is the so-called *linear module* (see e.g. [16], [26]). System $\Phi = (E_a; a = 1, 2, 3)$ of smooth vector fields on *B* tangent to *B* is a *vectorial basis* of W(*B*) (called also a *vector base* on *B*) if [23]:

(a) for each $p \in B$ the system $\Phi_p = (E_a(p); a = 1, 2, 3)$ is a base of the linear space $T_p(B)$ tangent to *B* at the point *p*.

(b) each field $\mathbf{v} \in W(B)$ is a linear combination fields of $\Phi$ with $C^\infty(B)$- coefficients.

If $\Phi$ is a vector base of W(B), then for each $\mathbf{v} \in W(B)$ the representation $\mathbf{v} = v^a \mathbf{E}_a$, $v^a \in C^\infty(B)$, is uniquely defined. A system $\Phi$ of smooth vector fields on B tangent to B is a vector base of W(B) if and only if the condition (a) is fulfilled [23].

We can define an internal operation in W(B) by means of the *Lie bracket*:

(A.7) $$[\mathbf{u}, \mathbf{v}] = \mathbf{u} \circ \mathbf{v} - \mathbf{v} \circ \mathbf{u}.$$

The Lie bracket is distributive with respect to addition and anticommutative; it is not associative but it satisfies instead the *Jacobi identity*:

(A.8) $$[\mathbf{u}_1, [\mathbf{u}_2, \mathbf{u}_3]] + [\mathbf{u}_2, [\mathbf{u}_3, \mathbf{u}_1]] + [\mathbf{u}_3, [\mathbf{u}_1, \mathbf{u}_2]] = 0.$$

The module W(B), together with the above-defined internal operation, constitutes a *Lie algebra* under the field $\mathbb{R}$ of real numbers [20]. Let us denote by $W^*(B)$ the linear module of smooth 1-forms on B. The 1-form $v = v_A dX^A \in W^*(B)$, $v_A \in C^\infty(B)$, is a $\mathbb{R}$-linear functional acting on fields $\mathbf{u} = u^A \partial_A \in W(B)$ according to the rule:

(A.9) $$\langle v, \mathbf{u} \rangle = v_A u^B \langle dX^A, \partial_B \rangle = v_A u^A, \qquad \langle dX^A, \partial_B \rangle = \delta_B^A.$$

However 1-forms are not $C^\infty(B)$-linear functionals. If $\Phi = (\mathbf{E}_a)$ is a vectorial basis of $W(B)$, then the vectorial basis $\Phi^* = (E^a)$ of $W^*(B)$ dual to $\Phi$ is univocally defined by the following condition:

(A.10) $$\mathbf{E}_a = e^A_a \partial_A, \quad E^a = \overset{a}{e}_A dX^A \quad \Rightarrow \quad \langle E^a, \mathbf{E}_b \rangle = \overset{a}{e}_A e^A_b = \delta_b^a.$$

If B is endowed with a Riemannian metric tensor **g**, then the smooth covariant and contravariant vector fields are in the following one-to-one correspondence:

(A.11) $W^*(B) \ni v = G(\mathbf{v})$, $\mathbf{v} \in W(B)$ iff $\forall \mathbf{u} \in W(B)$, $\langle v, \mathbf{u} \rangle = \mathbf{v}\mathbf{g}\mathbf{u}.$

In particular,

(A.12) $$\mathbf{E}^a = G(\mathbf{E}_a), \qquad \mathbf{E}^a \mathbf{g} \mathbf{E}_b = \delta^a_b.$$

Let us denote by $\pi = \{\pi_p \subset T_p B, \ p \in B\}$ a family of two-dimensional linear subspaces of spaces tangent to the three-dimensional Riemannian manifold $B_g = (B, \mathbf{g})$. The family defines the so-called two-dimensional *distribution* on $B_g$. Let $W_\pi(B)$ denote the set of all smooth vector fields on $B$ tangent to $B$ and identified with the linear first-order differential operators $\mathbf{u} = u^A \partial_A$, $u^A \in C^\infty(B)$, such that $\mathbf{u}_p \equiv \mathbf{u}(p) = u^A(X(p)) \partial_{A|p} \in \pi_p$ for each $p \in B$. Since $W_\pi(B)$ is closed under addition of these vector fields and their multiplication by real numbers as well as smooth functions on $B$, it is a *linear module* of smooth $\pi$-*fields* on $B$ [23]. The distribution $\pi$ is called *smooth* if for each point $p \in B$, there exists an open neighborhood $U$ of $p$ and if there are two linearly independent smooth vector fields $\mathbf{E}_1$, $\mathbf{E}_2$ tangent to $U$ such that $\mathbf{E}_1(q)$, $\mathbf{E}_2(q)$ span the linear space $\pi_q$, $q \in U$. The distribution $\pi$ is called *involutive* if for any $\mathbf{u}, \mathbf{v} \in W_\pi(B)$ we have $[\mathbf{u}, \mathbf{v}] \in W_\pi(B)$.

*Theorem* [23]

If $\pi$ is a *smooth* distribution, then the following conditions are equivalent:

(1) $\pi$ is an *involutive* distribution.

(2) There are $C^\infty(U)$-functions $C^\kappa_{\alpha\beta}$, $\alpha, \beta, \kappa = 1, 2$, such that

(A.13) $$[\mathbf{E}_\alpha, \mathbf{E}_\beta] = C^\kappa_{\alpha\beta} \mathbf{E}_\kappa.$$

(3) There exists a local vectorial basis $(\mathbf{c}_1, \mathbf{c}_2, \mathbf{c}_3)$ with a domain $U$ such that the fields $\mathbf{c}_1$ and $\mathbf{c}_2$ span the distribution $\pi$ on $U$ and

(A.14) $$[\mathbf{c}_1, \mathbf{c}_2] = 0.$$

(4) For each point $p$ there exists a map $(U, X)$, $p \in U$, such that fields $\mathbf{c}_A = \partial_A$, $A = 1, 2, 3$, constitute a local vectorial basis associated with this map (called its *natural base field*) and $\partial_1$, $\partial_2$ span the distribution $\pi$ on $U$.

Let $\mathbf{u} \in W(B)$ be a distinguished vector field. It is easy to see that we can introduce a $\mathbb{R}$-linear mapping $\delta: W(B) \to W(B)$ by means of the following formulae:

$$(A.15) \quad \begin{aligned} \delta &= \nabla^g_{\mathbf{u}} = u^A \nabla^g_A, & \mathbf{u} &= u^A \partial_A, \\ \delta \mathbf{v} &= \delta v^B \partial_B, & \delta v^B &= u^A \nabla^g_A v^B, & \mathbf{v} &= v^B \partial_B, \end{aligned}$$

where $\nabla^g$ denotes the Levi-Civita covariant derivative uniquely defined on the Riemannian space $B_g = (B, \mathbf{g})$ (a symmetric and metric with respect to $\mathbf{g}$ covariant derivative on $B_g$; e.g. [9], [20]). This linear mapping fulfils the so-called *Leibniz rule* with respect to multiplication of vector fields by scalars $f \in C^\infty(B)$:

$$(A.16) \quad \begin{aligned} \delta(f\mathbf{v}) &= f\delta(\mathbf{v}) + (\delta f)\mathbf{v}, \\ \delta f &= \nabla^g_{\mathbf{u}} f = \partial_{\mathbf{u}} f = u^A \partial_A f, \end{aligned}$$

and with respect to the scalar product defined on $B$ by the metric tensor $\mathbf{g}$:

$$(A.17) \quad \delta(\mathbf{v} \cdot \mathbf{w}) = \delta \mathbf{v} \cdot \mathbf{w} + \mathbf{v} \cdot \delta \mathbf{w}.$$

Such an operator $\delta$ is called a *variation operator* (in the direction of a dis-tinguished vector field $\mathbf{u} \in W(B)$). The value $\delta \mathbf{v}$ of this operator is called a *varia-tion* of the vector field $\mathbf{v}$ (in the direction of $\mathbf{u} \in W(B)$). Since

$$(A.18) \quad \begin{aligned} \delta v^A &= g^{AB} \delta v_B, & v_A &= g_{AB} v^B, \\ \delta v_A &= \nabla^g_{\mathbf{u}} v_A = u^B \nabla^g_B v_A, \end{aligned}$$

we have

$$(A.19) \quad \delta(\mathbf{u} \cdot \mathbf{u}) = 2\mathbf{u} \cdot \delta \mathbf{u} = 2 u^A \delta u_A = 2 u^A u^B \nabla^g_B u_A = 2 \varepsilon_{AB} u^A u^B,$$

where it was denoted:

(A.20) $$\varepsilon_{AB} = \frac{1}{2}\left(\nabla^g_A u_B + \nabla^g_B u_A\right).$$

The symmetric tensor field

(A.21) $$\boldsymbol{\varepsilon} = \varepsilon_{AB} dX^A \otimes dX^B,$$

is a Riemannian counterpart of the so-called *infinitesimally small strain* (or more briefly - *infinitesimal strain*) considered in the continuum mechanics.

Note that the formulae (A.17)-(A.19) are also valid if the covariant derivative $\nabla^g$ is replaced by an arbitrarily chosen covariant derivative $\nabla$. Moreover, in the literature is considered the so-called *infinitesimal variation operator* defined as (e.g. [9])

(A.22) $$\delta = dX^A \nabla_A,$$

and acting according to the rules $(A.17)_2$-(A.19) with $u^A$ replaced by $dX^A$. The value $\delta \mathbf{v}$ of this operator is called then an *infinitesimal variation* of the vector field $\mathbf{v}$ and interpreted as an *infinitesimal tangent vector field*. If the infinitesimal variation of a vector field vanishes:

(A.23) $$\delta \mathbf{v} = 0,$$

then it is said to be *covariant constant* over the differential manifold $(B, \nabla)$ [9].